\documentclass[aps,showpacs]{revtex4}
\usepackage{amsmath}
\usepackage{amsfonts}
\usepackage{amssymb,amscd}
\usepackage{epsfig}

\newcommand{\bea}{\begin{eqnarray}}
\newcommand{\eea}{\end{eqnarray}}
\newcommand{\spaceint}[2]{\int_{#1} d^3 #2 \;}
\newcommand{\vect}[1]{\mathbf{#1}}

\newcommand{\di}{\displaystyle}

\begin{document}

\title{Depletion force between two large spheres suspended in a bath of small spheres: 
 Onset of the Derjaguin limit}
\author{M. Oettel}
\affiliation{Max--Planck--Institut f\"ur Metallforschung, Heisenbergstr. 3, 70569 Stuttgart} 
\affiliation{Institut f\"ur Theoretische und Angewandte Physik, Universit\"at Stuttgart,
 Pfaffenwaldring 57, 70569 Stuttgart}

\date{\today}

\begin{abstract}
\rule{0ex}{3ex}
We analyze the depletion interaction between two hard colloids in a hard--sphere solvent
and pay special attention to the limit of large size ratio between colloids and solvent particles
which is governed by the well--known Derjaguin approximation. For separations between the colloids
of less than the diameter of the solvent particles (defining the depletion region),
the solvent structure between the colloids
can be analyzed in terms of an effective two--dimensional gas. Thereby we find that the Derjaguin
limit is approached more slowly than previously thought. This analysis is in good agreement with
simulation data which are available for a moderate size ratio of 10. Small discrepancies to
results from density functional theory (DFT) at this size ratio become amplified for
larger size ratios.
Therefore we have improved upon previous DFT techniques by imposing test particle consistency
which connects DFT to integral equations. However, the improved results show no convergence
towards the Derjaguin limit and thus we conclude that this implementation of DFT together with
previous ones which rely on test particle insertion become
unreliable in predicting the force in the depletion region for size ratios larger than 10.
\end{abstract}


\maketitle

\section{Introduction and physical problem}

In many circumstances the effective interaction between larger colloidal particles suspended
in a bath of smaller particles is determined by entropic effects. 
If both colloidal and solvent particle are taken to be hard spheres with radii $R_2$
and $R_1=\sigma /2$, respectively, the colloidal interaction is purely entropic and arises
mainly through the effect of a depletion zone between the colloids (for surface--to--surface 
minimal distances $z<\sigma$) which is inaccessible to the solvent particles.

For large size ratios $\alpha=R_2/R_1$, the  force $F_\alpha$ in the depletion zone $z<\sigma$ 
between the colloids can be obtained by just using bulk and surface thermodynamics. This
is the Derjaguin approximation (its derivation is presented in more detail below) which 
states:
\bea
 \frac{F_\alpha(z)}{\pi(R_1+R_2)} =  p(z-\sigma) - 2\gamma_\infty\;, \qquad z \le \sigma\;.
\eea
Here, $p$ is the bulk pressure at density $\rho$ of the small spheres and $\gamma_\infty$ is
the surface tension for small spheres of density $\rho$ at a planar hard wall. For both quantities
quasi--exact expressions are available \cite{Car69,Hen87},
\bea
 \label{p3d}
 \frac{ p }{\rho} &=& \frac{1+\eta+\eta^2-\eta^3}{(1-\eta)^3} \;, \\
 \label{g3d}
  \frac{\gamma_\infty }{\rho} &=& -\frac{\frac{3}{4}\eta(1+\frac{44}{35}\eta-\frac{4}{5}\eta^2)}
           {(1-\eta)^3} \;, \qquad \eta=\frac{\pi}{6}\;\rho\;.
\eea
In obtaining these equations we have set
\bea
   \beta=\sigma=1
\eea
and we will do likewise in all following considerations.

\begin{figure}
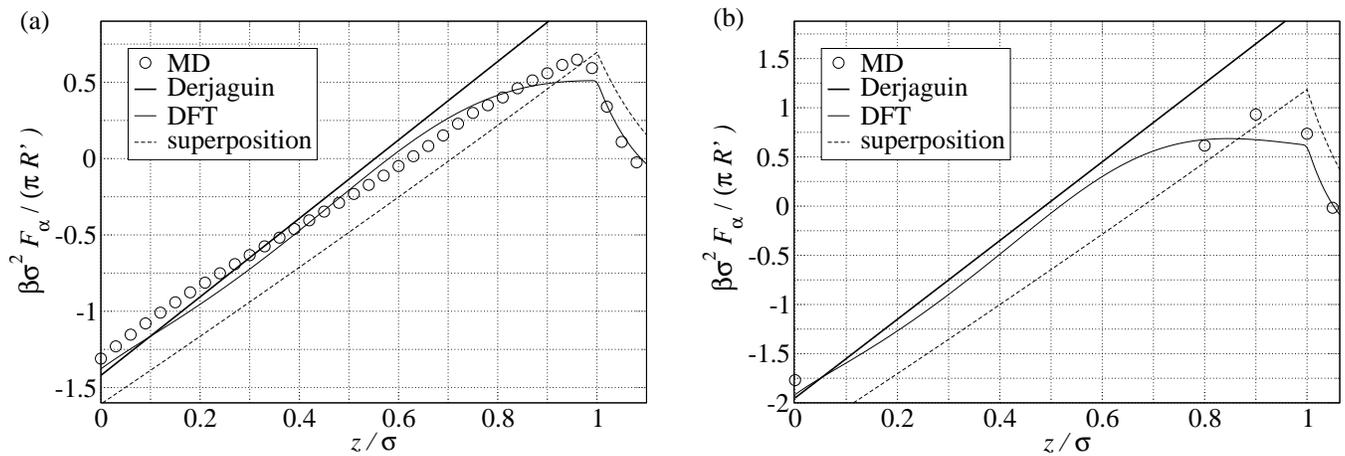

 \begin{center}
   \epsfig{file=fig1a.eps,width=8.5cm} \hspace{5mm} 
   \epsfig{file=fig1b.eps,width=8.5cm} 
 \end{center}
 \caption{A comparison between existing results for the force between two colloidal particles 
  in the depletion region for the size ratio $\alpha=10$. 
  Shown are results from molecular dynamics 
  \cite{Bib96}, DFT \cite{Rot00},
  the superposition approximation (using density profiles obtained as in \cite{Rot00}), and the
  Derjaguin limit for solvent densities: (a) $\rho=0.6$ and (b) $\rho=0.7$.}
 \label{compar-fig}
\end{figure}

In recent years, progress has been made in calculating the depletion force in hard systems
by other theoretical means, such as simulations \cite{Bib96,Dic97}, integral equations \cite{Dic97} 
and density functional theory (DFT) \cite{Rot00}.
In a recent paper \cite{Hen02}, Henderson reviews this analysis of depletion forces 
in hard fluids and points to a serious discrepancy between Derjaguin's analysis on the
one hand, and the various theoretical approaches/molecular dynamics simulations on the other hand.
Also there are some features of the density functional result which do not fit with the simulations
either. These discrepancies had been gone unnoticed partly due to the fact that comparisons
were made between depletion {\em potentials} which add a fair amount of uncertainty to the
simulation data since force curves with very few data points had to be integrated \cite{Bib96}.
Also the limits inherent in the Derjaguin assumption have not been analyzed convincingly
such that partial agreement respectively disagreement with Derjaguin's result has not been
taken seriously. Henderson's analysis concentrated on the depletion force between hard walls and
hard colloids but applies equally well to the force between two colloids.

Before we analyze these discrepancies, we briefly present the strategies of the various
approaches to obtain the depletion potential. Let us denote by 
$\rho(\vect r;\; \vect x_1, \vect x_2)$ the density distribution of small spheres
 around two fixed hard spheres at positions $\vect x_1$ and $\vect x_2$. Then the depletion 
force on one big sphere can be obtained by summing over all small spheres the force between 
a single small sphere and the big sphere. By symmetry, the force will be directed along the
axis joining the centers of the two big spheres and due to the hard sphere interactions
the volume integral  reduces to an integral over the surface of one big sphere. Its magnitude
(negative for attraction, positive for repulsion) is given by
\bea
  \label{F-mc}
  F_\alpha(z) &=& 2\pi (R_1+R_2)^2\;\int_{-1}^1 d(\cos\theta) \cos\theta\;
    \rho(\vect r;\; \vect 0, \vect x_2) \;, \\
   && \left[ |\vect r|=R_1+R_2\;, \quad  \vect x_2=(R_2+z,0,0)\;\right] \;. \nonumber
\eea
In simulations, just this formula is used. The superposition approximation also uses this
formula and additionally assumes
that $\rho(\vect r;\; \vect x_1, \vect x_2)$ can be obtained by superimposing the
two density distributions $\rho(\vect r- \vect x_i)$ around one fixed hard sphere centered at
$\vect x_1$ and $\vect x_2$, respectively:
\bea
  \label{superpos}
  \rho(\vect r;\; \vect x_1, \vect x_2) = \frac{1}{\rho}\; \rho(\vect r- \vect x_1)
  \rho(\vect r- \vect x_2)\;.
\eea
The density distribution around one big sphere could be determined by e.g. integral equation 
methods or by minimizing a density functional. When presenting superposition approximation 
results, we will use DFT results using the Rosenfeld functional as these are of superior 
quality. The DFT method of Ref.~\cite{Rot00} also arrives at the depletion potential (whose
derivative gives the depletion force) by just using the density distribution around one single
big sphere but circumvents the crude approximation, eq.~(\ref{superpos}), by making use
of the potential distribution theorem (also known as Widom's insertion trick). 
The method (which we call insertion route DFT) is explained in App.~\ref{dft-app}. 
On the other hand, the distribution $\rho(\vect r;\; \vect x_1, \vect x_2)$ could be
obtained directly using DFT (in line with Ref.~\cite{Rot00} we call this brute force DFT).
It is numerically involved and only two studies exist in the literature, both for
size ratios smaller or equal  $\alpha = 5$ \cite{Mel00,Gou01}. Error bars on the results of
Ref.~\cite{Mel00} are much too large to arrive at a sensible conclusion. The much improved
results of Ref.~\cite{Gou01} indicate no significant deviation between the depletion potentials
calculated using the insertion route and the brute force method, respectively.

For densities $\rho>0.5$ discrepancies between the above mentioned treatments and the simple Derjaguin
formula become apparent as is illustrated in Fig.~\ref{compar-fig}. For $\alpha=10$ and solvent
sphere densities $\rho=0.6$ and 0.7 we show MD data \cite{Bib96}, Derjaguin's result, 
insertion route DFT data calculated as in Ref.~\cite{Rot00} 
and data obtained from the superposition approximation.
The deviation from Derjaguin's straight line is most obvious near $z=1$, i.e. near where just one
small sphere fits between the two large spheres. 
The MD results seem to follow a straight line
with a slope smaller than the one in Derjaguin's expression, $p$, but with a characteristic rounding
off near $z=1$ which always overshoots the DFT data (see also Fig.~7 in Ref.~\cite{Dic97} 
for another simulation). 
The DFT results show a flattening off which is characteristic for $\rho>0.5$ and $\alpha>10$. The
same behavior is seen in results using bridge diagram corrected HNC integral equations
\cite{Dic97}. Finally, the superposition approximation produces a straight line with Derjaguin's 
slope but with a big offset. Using density distributions $\rho(\vect r- \vect R)$ from less precise
methods (integral equations with Percus--Yevick and Rogers--Young closure) offset and slope
of the straight line are changed considerably \cite{Bib96} such that these results fitted
the MD data quite well. This led the authors of \cite{Bib96} to the erroneous conclusion
that the superposition approximation is quite succesful in predicting 
$\rho(\vect r;\; \vect x_1, \vect x_2)$. From the present results, it is clear that
the superposition approximation does not constitute a good model of the force in the depletion zone.

At first glance one is inclined to blame the discrepancies on the finiteness of $\alpha$. After all,
Derjaguin's result is supposed to be valid for $\alpha\to \infty$.
Here the first problem arises:
regarding this limit, Henderson gives
an argument (which will be critically examined below) that deviations to Derjaguin
should only occur for $z> 1-1/(4R_1+4R_2)$ ($z>0.955$ for $\alpha=10$). 
This is clearly not the case for
all results as can be seen in Fig.~\ref{compar-fig}. 
Moreover the insertion route DFT results do not converge to the Derjaguin limit for higher $\alpha$ 
\cite{Rot00}.
The second problem lies in the fact that Rosenfeld's (or related) DFT usually gives 
density distributions around fixed objects (wall \cite{Kie90}, 
big spheres \cite{Rot02}, wedge \cite{Bry03})
of such a high quality that they seem to parametrize MC/MD data also for higher densities.
However, in the present case systematic discrepancies between the MD and the DFT results
occur. A tentative first explanation why this happens lies in the possibility 
that insertion route and brute force DFT give substantially different results for $\alpha \ge 10$
(remember, there is no apparent difference for $\alpha = 5$ \cite{Gou01}). A second
possibility is that the higher--order correlations which are captured only approximately by any DFT
model become more and more important. In fact, we will present below a picture for the depletion
force which reveals quite subtle packing effects between the colloids which emerge for larger
values of $\alpha$.

There is an interesting consequence from all of this. Defining the depletion potential by
\bea
 \label{wdef}
 W_\alpha(z) = \int_z^\infty F_\alpha(z')\; dz'\; ,
\eea
we note that in the Derjaguin approximation $W_\alpha(0) < W_\alpha(1)$ only for $\rho< 0.68$. Above
that density, contact between the two big spheres is only a metastable minimum separated
by a rather high potential barrier from the overall minimum which will be close to $z=1$.
So, for higher densities the colloidal particles would not stick to each other.
Although according to insertion route DFT $W_\alpha(0) -W_\alpha(1)$ also increases with 
increasing $\rho>0.7$,
this quantity never changes its negative sign for physical densities\footnote{We checked 
this for $\alpha\le 100$.}.  

Therefore we can formulate our questions: Does the Derjaguin limit already set in
for $\alpha \approx 10$? If not, why? What is the source of discrepancy between DFT and MD/MC?
As Rosenfeld's DFT is now being used in studies of solvation forces for 
colloidal particles in liquids with interactions other than hard sphere \cite{Amo01},
the understanding of its limits for hard spheres is crucial.

The remainder of the paper is organized as follows. In order to have a self--contained
presentation, the Derjaguin limit for the depletion force is derived via $(i)$ a force and 
$(ii)$ energy analysis
and $(iii)$ exact relations from statistical mechanics. This section contains nothing new and leans
heavily on the presentation in Ref.~\cite{Hen02}.
To shed light on the onset of the Derjaguin limit,
 we will rederive it in a slightly different way in the following section and thus show
that it is not valid when the colloids are separated by $z \approx 1$. 
This will define the new {\em annular slit approximation}.
The depletion regime $z<1$ is then analyzed in terms of an effective two--dimensional system of small 
disks which builts
up in the annular wedge between the colloids. 
Using scaled particle theory in two dimensions, we derive an expression for $F_\alpha(z)$
which for $\alpha\to\infty$ recovers the Derjaguin expression, although at a smaller rate
as Henderson anticipated. 

 For $\alpha=10$, the results of this analysis point to a flaw in the
insertion route DFT treatment and show better agreement with the MD data.
Therefore we will examine the insertion route DFT results closer and improve upon them 
by imposing test particle consistency (see App.~\ref{dft-app}). The equations obtained
can also be viewed as RHNC integral equations with the bridge diagrams calculated from the 
hard--sphere density functional. Therefore, results from any integral equation closure
can be viewed as being akin to insertion route DFT calculations. The quality of the density 
functional is then closely related to the quality of the bridge function approximation. 

Using test particle consistent DFT, we find no convergence to Derjaguin's result for 
$\alpha$ up to 100 and an increasing
difference to the annular slit approximation (which becomes more reliable for increasing $\rho$ and
$\alpha$). With the premise that the Dejaguin limit is reached in a non--singular way,
we arrive therefore at the conclusion that insertion route DFT (and likewise integral equations) 
are unreliable for $\alpha > 10$ since they miss some of  the packing effects of the small spheres
between the large colloids. 

In the last section we briefly comment on the possibility that
non--analytic contributions might prevent a smooth transition to the Derjaguin limit.

\section{Derjaguin approximation}

\subsection{Force analysis}

\begin{figure}
 \begin{center}
   \epsfig{file=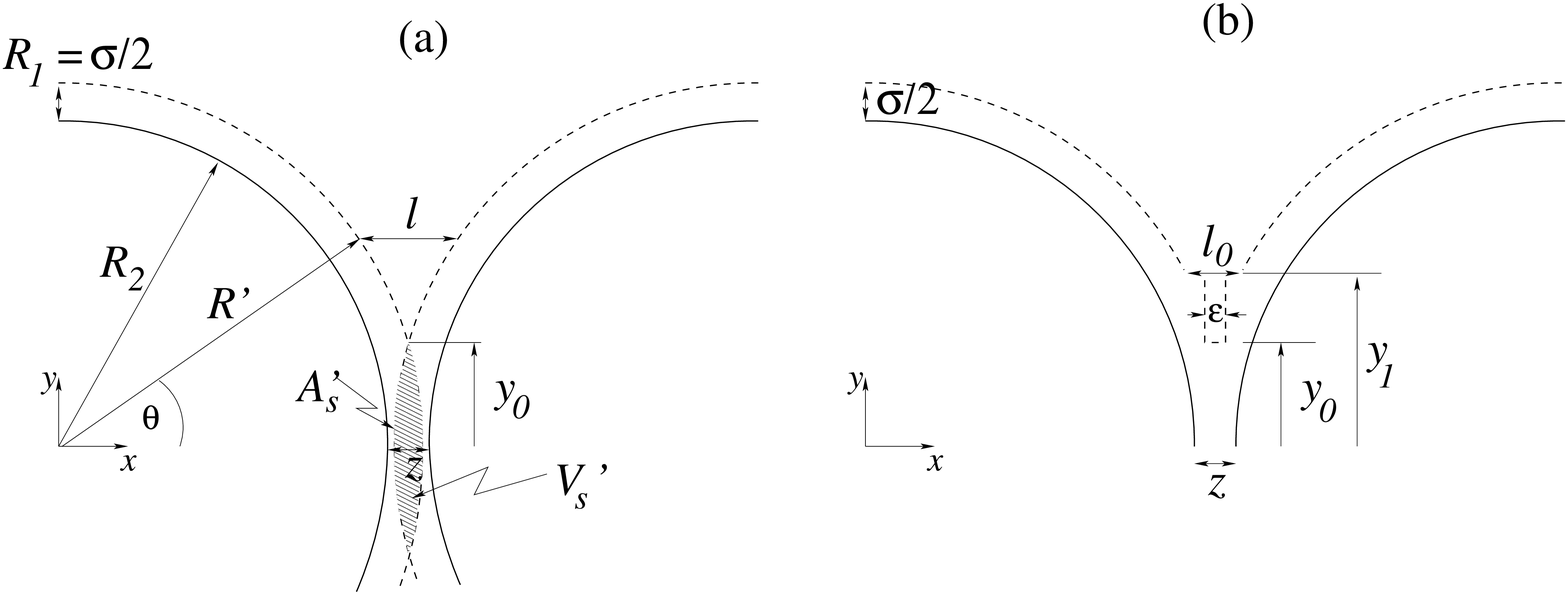, width=17.5cm}
 \end{center}
 \caption{(a) Geometrical definitions for two colloids in the depletion regime.
  The dashed lines indicate the surfaces of the exclusion spheres. Their overlap defines
  a volume $V_s'$ and an overlap surface area $A_s'$. (b) Modelling
  the annular wedge part with widths $l<l_0$ by an annular slit of width $\epsilon$
  in which the solvent gas is effectively two--dimensional. }
 \label{sph1-fig}
\end{figure}

The geometrical arrangement of the two colloids is shown in Fig.~\ref{sph1-fig}.
The boundary of the exclusion zone for
the centers of the small particles is indicated by the dashed lines, thus the exclusion
zone are two (possibly overlapping) spheres of radius $R'=R_1+R_2$.
The depletion force between the two large spheres is obtained by summing local pressures over
the area of one (exclusion) sphere,
\bea
 \label{F1}
   F_\alpha (z) = 2\pi R'^2 \int_{-1}^1 d(\cos\theta) p_{\rm loc}(\theta) \;.
\eea  
The Derjaguin approximation consists in replacing the local pressure by the solvation
force per unit area of a planar slit with width $l$ where the width refers to the minimal 
distance between the excluded volumes of the walls: 
\bea
   p_{\rm loc}(\theta) \approx f_\infty (l) \;.
\eea
Here $l$ is the horizontal distance between the two (exclusion) spheres corresponding 
to the angle $\theta$,
see Fig.~\ref{sph1-fig}. For very large $\alpha$, this approximation is certainly justified, since 
locally the geometry resembles the planar slit. Using
\bea
  \label{co1}
  l &=& 2R'-2x-(1-z) \;, \\
  \label{co2}
  \cos \theta &=& x/R' \;,
\eea
we transform eq.~(\ref{F1}) into
\bea
  \label{Derjag1}
  F_\alpha (z) = \pi R' \int_{z-1}^\infty dl f_\infty(l)
\eea  
The upper limit in this integral over the slit width has been replaced by infinity since
it can be assumed that the solvation force approaches its limiting value $f_\infty(l \to \infty)=0$
when $l$ is just a few $\sigma$,  which should be considerably smaller than $R'$.
Now the solvation force per unit area is defined by
\bea
 f_\infty(l) = - \frac{ d\gamma(l)}{dl}\;,
\eea
where $\gamma(l)$ is the excess grand potential (i.e. bulk grand potential subtracted) 
of the system of the two parallel walls which define
the slit. Using the fact that $\gamma(\infty) = 2\gamma_\infty$, where $\gamma_\infty$ is the surface
tension of a single hard wall in a sea of small spheres we find
\bea
  F_\alpha (z) = \pi R'( \gamma(z-1)-2\gamma_\infty ) \;.
\eea
If $z<1$ (i.e. no single small sphere fits into the slit) the surface tension  
arises from the release of free volume to the small spheres, $\gamma = p(z-1)$. Thus
\bea
  \label{derjag1}
 F_\alpha(z) = \pi R'( p(z-1) - 2\gamma_\infty)\;, \qquad z \le 1\;.
\eea
The depletion force is seen to depend only on the (hard sphere) pressure $p$ and surface tension
$\gamma_\infty$ for which we possess accurate approximations, see eqs.~(\ref{p3d},\ref{g3d}).

%
% This figure should be wide (two columns) !
%

\subsection{Energy analysis}

Following Henderson \cite{Hen02} we can arrive at the Derjaguin approximation also by an analysis
of the grand potential which can be decomposed into a 'volume', a 'surface area' and a 'line'
term according to
\bea
 \Omega(z) &=& -p\;V_s(z) + 2\pi R'^2 \int_{-1}^1 d(\cos\theta) \gamma(l) \; \\
  & = & -p\;V_s(z) + \gamma_\infty \; A_s(z) + \pi R' \int_0^\infty dl 
   (\gamma(l)-2\gamma_\infty)\;.
\eea 
Here, $V_s$ is the volume available to the small spheres (i.e. outside the two (possibly overlapping)
exclusion spheres) and  $A_s$ is the corresponding surface area of the two (possibly overlapping) 
exclusion spheres,
\bea
  V_s(z) & = & V_0-V_s'= V_0 - \frac{8}{3}\pi R'^3 +\frac{\pi}{2}(z-1)^2\left(R'+\frac{z-1}{6}\right)\; \\
  A_s(z) &=& A_0-A_s' =8\pi R'^2+  2\pi R'(z-1) \; .
\eea
($V_0$ is the total system volume)
 and the last term is the 'line tension' contribution {\em independent} of
the length of the overlap circle $2\pi y_0$ and thus {\em independent} of $z$. 
Using 
\bea
 \label{fdep-def}
 F_\alpha(z) = - \frac{\partial \Omega}{\partial z} 
\eea
and retaining only the leading terms in $1/R'$, one arrives at the Derjaguin result, 
eq.~(\ref{derjag1}).

\subsection{Statistical mechanical analysis}

Summarizing Henderson's analysis, let us consider a fixed big sphere  surrounded by a bath of small
spheres. This sphere exerts  an external, hard--body potential
$V^{\rm ext}_i$ on small spheres [$i=1$] and on any other large spheres [$i=2$]). 
The work to insert another big sphere at distance $z$ from the first sphere
is given by $- c^{(1)}_2(z)$. Here, $c^{(1)}_2(z)$ is the one--body correlation
function given by
\bea
  c^{(1)}_2(z) &=& \log(\rho_2(z)\Lambda_2^3) -\tilde\mu_2(z)\;,  \\
  \tilde\mu_2(z) &=& \mu_2 -V^{\rm ext}_2(z)  \; .
\eea
In these expressions, $\Lambda_2$ is the de-Broglie--wavelength of the big spheres and $\mu_2$ is
their chemical potential.
For our configuration of interest, $\rho_2(z)$ is the one--body density profile of big spheres on
another big sphere in the bath of small spheres {\em in the dilute limit} ($\mu_2 \to -\infty$). 

The first two equations in a hierarchy of functional derivatives of the grand potential are
\bea
  \frac{\delta \Omega}{\delta \tilde\mu_i(\vect x)} &=& -\rho_i(\vect x) \; , \\
   \frac{\delta^2\Omega}{\delta \tilde\mu_i(\vect x)\delta \tilde\mu_j(\vect y)} &=& -\left[
      \rho_i(\vect x)\rho_j(\vect y)(g_{ij}(\vect x,\vect y)-1) + 
     \rho_i(\vect x)\delta^{(3)}(\vect x-\vect y)\delta_{ij}\right]\; .
\eea
Let us assume that the center of the first, fixed sphere defines the origin of the coordinate
system, and the coordinates of the center of the  second big sphere are given by 
$\vect x=\{2R_2+z,0,0\}$.
Since the external potential $V_2$ vanishes for $z>0$ we find,
using the above  equations and the definition of the depletion force, eq.~(\ref{fdep-def}), 
\bea
 F_\alpha(z) = c^{(1)\prime}_2(z)&=& \frac{\rho_2'(z)}{\rho_2(z)}  \qquad (z>0) \;, \\
                &=& \frac{1}{\rho_2(z)} \int d^3\vect y \sum_{i,j=1,2} \frac{\delta^2\Omega}
                {\delta \tilde\mu_i(\vect x)\delta \tilde\mu_j(\vect y)} \; 
    \frac{\partial \tilde\mu_j(\vect y)}
                {\partial y_1} \; .
\eea
In the dilute limit $(\rho_2\to 0)$ this expression simplifies to
\bea
  F_\alpha(z) &=& - \int d^3\vect y  \frac{\partial V_1^{\rm ext}(\vect y)} {\partial y_1} \;
    \rho_1(\vect y)( g_{12}(\vect x,\vect y)  -1 ) \; .
\eea
By virtue of the derivative of the external potential $V_1^{\rm ext}$ (exerted by the fixed sphere
on the small spheres) only the surface of the exclusion sphere (i.e. one of the dashed lines
in Fig.~\ref{sph1-fig}) contributes to the integral. Since $\rho_1(\vect y)$ is the density profile
of small spheres around the fixed sphere, and $\rho_1(\vect y) g_{12}(\vect x,\vect y)=
\rho(\vect y;\;\vect 0,\vect x)$ 
is the density profile
of small spheres around the fixed sphere but with the second sphere fixed at position $\vect x$,
we recover eq.~(\ref{F-mc}):
\bea
  \label{d2}
   F_\alpha(z) &=& 2\pi R'^2 \int_{-1}^1 d(\cos\theta) \cos\theta 
    \;\left(\rho(\{R',\theta\};\vect 0,\vect x)-\rho_1(R') \right)\; , \\
         &=& 2\pi R'^2 \int_{-1}^{\cos\theta_{\rm min}} d(\cos\theta) \cos\theta
    \;\rho(\{R',\theta\};\vect 0, \vect x) \; ,\\
    && \cos\theta_{\rm min} =\left\{ \begin{matrix} 1 - (1-z)/(2R') & (z<1) \\
                  1 & (z>1) \end{matrix} \right.
\eea
Although the integral in eq.~(\ref{F-mc}) extends over the whole surface of the big sphere,
the contribution of surface elements with azimuthal angle $\theta<\theta_{\rm min}$ is zero
since the density vanishes there.

To identify the Derjaguin limit it is useful to keep the second term
in the brackets on the rhs of eq.~(\ref{d2})  (writing for the contact density of small
spheres at a single large sphere $\rho_s=\rho_1(R')$): 
\bea
 F_\alpha(z<1) &=& 2\pi R'^2 \int_{-1}^{\cos\theta_{\rm min}} a\;da 
    \;(\rho(\{R',a\};\vect 0,\vect x)-\rho_s) + 
 2\pi R'^2\rho_s \int_{-1}^{\cos\theta_{\rm min}}a\; da \\
    &=&  2\pi R'^2 \int_{-1}^{\cos\theta_{\rm min}} a\;da
    \;(\rho(\{R',a\};\vect 0,\vect x)-\rho_s) + \pi R' \rho_s (z-1)\left(1+\frac{z-1}{4R'}\right)\; .
\eea
Changing integration variables from $a=\cos\theta$ to $l$ according to eqs.~(\ref{co1},\ref{co2}) and
identifying the upper limit with infinity, we find
\bea
  F_\alpha(z<1) &=&  \pi \int_0^\infty dl (R'-(l+1-z))\;(\rho(\{R',l\};\vect 0,\vect x) - \rho_s) +
      \pi R' \rho_s (z-1)\left(1+\frac{z-1}{4R'}\right)\; .
  \label{d1}
\eea
The ``microscopic'' Derjaguin approximation consists in setting
\bea
 \label{Derjag_mic}
 \cos\theta \left(\rho(\{R',l\};\vect 0,\vect x) - \rho_s\right) = \left( 1- \frac{l+1-z}{2R'}\right)
  \left(\rho(\{R',l\};\vect 0,\vect x) - \rho_s\right) \approx \rho_w(l)-\rho_w \;,
\eea
where $\rho_w(l)$ is the contact density at one wall in a planar slit of width $l$ and $\rho_w$
is the contact density at a single planar wall. From statistical mechanics we furthermore know
\cite{Hen86,Hen86_2}
\bea
  \rho_w(l)-\rho_w &=& f_\infty(l) = -\frac{d \gamma(l)}{dl} \;, \\
    \rho_s &=& p + \frac{2\gamma_{R'}}{R'} + \frac{d\gamma_{R'}}{dR'} \; .
\eea
Putting the last two equations into eq.~(\ref{d1}) we find the Derjaguin result as the leading 
order in $R'$, and we can identify the finite--size correction of first order to it
(the surface tension on the (exclusion) sphere with radius $R'$ can be approximated in first order
by the surface tension on a hard wall, $\gamma_{R'}\approx\gamma_\infty$):
\bea
 \label{derjag-sm}
 F_\alpha(z<1) & \approx & F_\alpha^{\rm Derjag} + \pi R'\; \frac{2\gamma_\infty+\frac{z-1}{4}}{R'}(z-1)\; .
\eea
Interestingly, the finite size corrections predict a smaller slope for the force curves
($\approx 10$\% for $\alpha=10$) and a slight deviation from linearity which affects the curve
only for $z\to 0$. We note that the considerations of Ref.~\cite{Goe98} (their 
{\em wedge approximation}) 
would modify our finite--size corrections by $ F_\alpha/(\pi R') \to  F_\alpha/(\pi R') -
\gamma_\infty/R'$, i.e. the slope corrections would be mitigated. In any case, 
the qualitative behaviour for $z\to 1$ remains unchanged.

\section{ Annular slit approximation and Derjaguin limit}

The microscopic Derjaguin approximation of eq.~(\ref{Derjag_mic}) asserts that -- apart from
the geometrical factor $\cos\theta$ -- all annular wedges that are formed between the
two large spheres for $z<1$ are equivalent, i.e. the contact value of the density on the spheres
can be described by a single function, namely $\rho_w(l)$.
At first glance, there is a physical difference between these wedges: 
At $z=1$ the spheres on one ``side'' of the annular wedge can 
scatter with the spheres on the other side , as opposed to smaller values of $z$.
Henderson argues that 
for small values of $l$, an effectively
two--dimensional {\em ideal} gas of small spheres forms between the two colloids since
the limiting three-dimensional density $\rho_w(l\to 0)$ stays {\em finite} and therefore an 
effective 2D density $\rho_{\rm 2d}\approx l\; \rho_w$ {\em vanishes}. Therefore, scattering
from one side of the wedge to the other should be negligible {\em unless} zero separation
between the colloids occurs for radial distances $y_0<1/2$. Since $y_0^2 \approx (1-z)R'$,
it follows that the Derjaguin approximation should be valid for $z<1-1/(4R')$.   

Is the concept of a nearly ideal 2D gas really valid in the annular wedge?
For narrow slits with finite $l$ we  consider
the small $l \ll \sigma $ expansion of $\rho_w(l)$ \cite{Hen86}:
\bea
 \label{small_l}
 \rho_w(l) &=& \frac{1}{  \rho^{-1} \exp( -\mu^{\rm ex}) - \pi\;l}\;. 
\eea 
Here, $\mu^{\rm ex}$ is the excess chemical potential of the small spheres and $\rho$ the 
corresponding density. However, this limiting behavior is valid only for {\em very} small $l$.
Rather one should study the effective 2D density in a slit defined by
\bea
 \label{coverage}
  \rho^{\rm slit}_{\rm 2d}(l) = \int_0^l dl' \rho(l') = \rho_0\; l + \Gamma(l) \; ,
\eea
with $\Gamma(l)$ defining the coverage in the slit. A Derjaguin--like estimate 
for the average
2D density in the annular wedge up to a maximal parallel distance $l_0$ of the exclusion spheres
follows:
\bea
 \label{rho_av}
  \rho_{\rm 2d}^{\rm av}(l_0) = \frac{1}{l_0} \int_0^{l_0} \rho^{\rm slit}_{\rm 2d}(l)=
  \frac{1}{l_0}\int_0^{l_0} \Gamma(l)dl+ \frac{1}{2}\rho_0 l_0 \;.
\eea 

\begin{figure}
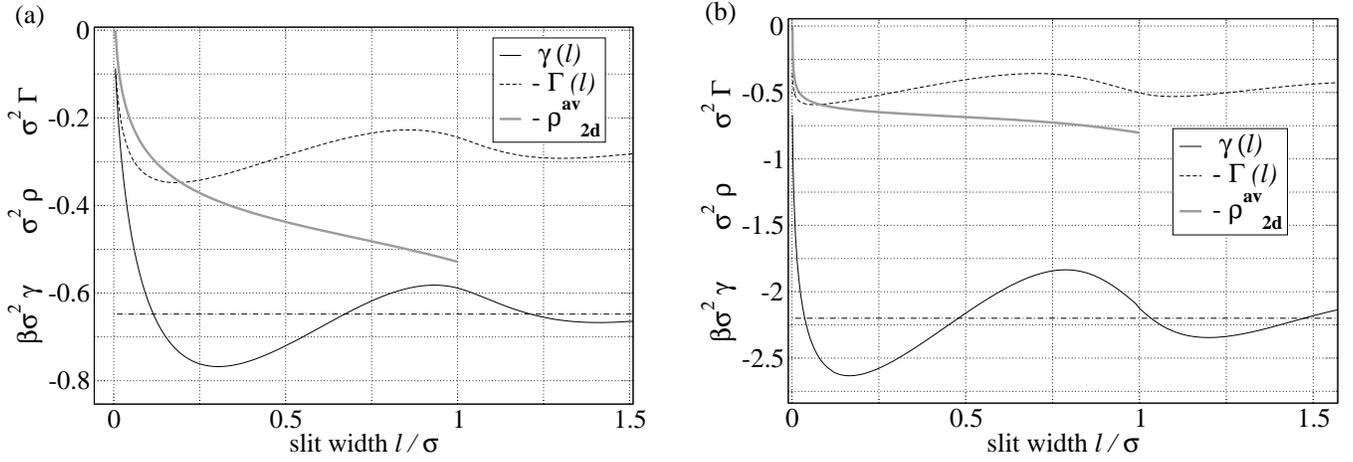

 \begin{center}
   \epsfig{file=fig3a.eps,width=8.5cm} \hspace{5mm}
   \epsfig{file=fig3b.eps,width=8.5cm} 
 \end{center}
 \caption{Surface tension $\gamma(l)$ and the negative of the coverage $\Gamma(l)$ 
 for planar slits of width $l$ and
 two solvent densities: (a) $\rho=0.5$ and (b) $\rho = 0.7$, as obtained from minimizing the Rosenfeld functional. 
 The dot--dashed line shows the surface tension
 $\gamma(l\to\infty)=2\gamma_\infty$. The (negative of the) average density refers to the 
  Derjaguin like
 approximation of the 2D density in the annular wedge, eq.~(\ref{rho_av}).}
 \label{slit-fig}
\end{figure}

We can gain access to this quantity by using DFT again. As explained earlier, minimizing the
Rosenfeld functional in the presence of an external field gives rather accurate density 
distributions. Carrying out the minimization in the presence of the two hard walls which define
the slit gives us the explicit density distribution in the slit from which the surface tension 
$\gamma$, the coverage and the average 2D density as functions of $l$ can be calculated. The results 
for two medium densities are shown in Fig.~\ref{slit-fig}.
It is seen that the coverage and therefore also the average 2D density quickly reaches the level
of $2\:\Gamma_\infty$ (twice the coverage on a single planar wall) and therefore the 2D gas between
the spheres is far from ideal. We also notice that the surface tension $\gamma(l)$ falls quickly 
to $2\gamma_\infty$ and then shows moderate oscillations around that value.

Now, in order to formuate an alternative derivation of the Derjaguin limit we replace the last
 part of the 
annular wedge with $l<l_0$ by an annular slit of width 1+$\epsilon$ where the spheres can only
move perpendicular to the $z$--axis, see the right 
panel of Fig.~\ref{sph1-fig}. The 
spheres in the slit can then be viewed as a system of hard disks.
The surface tension in this
fictitious slit is $\gamma(l_0)$ and its  surface grand potential is written as
\bea
 \label{om_sur}
 \Omega_{\rm sur} &=& \gamma(l_0) \; A_{\rm wed} + \sigma(y_0)\; 2\pi y_0 + \dots \; .
\eea
where $A_{\rm wed}$ is the one--sided area of the wedge and
we have introduced a line tension term $\sigma(y_0)$ which describes the interaction at the
inner boundary of the wedge. Now for large $R'$ we have
\bea
 \label{geom}
  y_0 \approx \sqrt{R'(1-z)}\;, \qquad A_{\rm wed} \approx A_0(y_1) +\pi R'(z-1) \;,
\eea
where the area of a spherical cap is $A_0 \approx \pi y_1^2$. Now the depletion force
has three contributions:
\bea
  \label{e1}
  F_\alpha (z<1) &=& \left(p+\frac{2\gamma_\infty}{R'}\right)\;\frac{dV_s}{dz} - \frac{d\Omega_{\rm sur}}{dz} + \int_{l_0}^\infty f_\infty(l)dl \;,
\eea
which arise since we have split the original Derjaguin integral, eq.~(\ref{Derjag1}), 
according to $\int_{z-1}^\infty \dots = \int_{z-1}^{0}\dots+\int_0^{l_0}\dots+\int_{l_0}^\infty\dots$
and have incoroprated the finite--size correction of eq.~(\ref{derjag-sm}).
Simplifying eq.~(\ref{e1}) using the geometrical relations in eq.~(\ref{geom}) we find,
\bea
 \label{e2}
 F_\alpha (z<1) &=& \pi R'\left[ \left(p+\frac{2\gamma_\infty}{R'}\right)\left( (z-1) + \frac{(z-1)^2}{4R'}
   \right) - \gamma(l_0) + \frac{\sigma(y_0)}{y_0}
    + \sigma'(y_0) - (2\gamma_\infty - \gamma(l_0)) \right].
\eea
We have recovered the finite--size corrected Derjaguin result, eq.~(\ref{Derjag1}), 
plus some line tension contribution which allows
us to formulate Henderson's hypothesis of the equivalence of the annular wedges as
\bea
 \lim_{y_0\to 0} \frac{\sigma(y_0)}{y_0} = \lim_{y_0\to 0} \sigma'(y_0) = 0 \; .
\eea 
This is not trivial at all. Rather, if $\sigma(y_0)$ or $\sigma'(y_0)$ go to a finite value
as $y_0\to 0$ we would expect the Derjaguin limit to fail for $z \to 1$.

\subsection{ The effective 2D gas in the annular slit} 

We can gain access to the line tension function $\sigma(y_0)$ by exploiting the 
nature of the quasi--2D gas in the wedge.
Since the small spheres can only move in the plane perpendicular to the axis joining the centers
of the two colloids, we rewrite the surface grand potential of our fictitious slit as 
\bea
  \Omega_{\rm sur} \to \Omega_{\rm 2d} = -p_{\rm 2d}(\rho_{\rm 2d}) A_{\rm wed} +
  \gamma_{\rm 2d}(\rho_{\rm 2d}; y_0) \; 2\pi y_0  \; ,
\eea
appropriate for the 2D ``volume'' and ``area'' grand potential contributions of a system of 
hard disks.
Now let us think about what is physically happening when the second colloid approaches the
first one at distances $z\le 1$: A circular cavity forms in the center of the quasi--2D gas
which cannot be reached by the centers of the solvent spheres.
Therefore we can express the last equation as
\bea
  \Omega_{\rm 2d} = -p_{\rm 2d}\;A_0(y_1) + \mu_{\rm cav}(y_0')
\eea 
Here, the first term is the $z$--independent 2D ``volume'' term ($A_0(y_1) \approx \pi y_1^2$,
see the right panel of Fig.~\ref{sph1-fig}),
and we have introduced
 $\mu_{\rm cav}(y_0')$, the work needed to create a cavity of radius $y_0'$.
From the reasoning above we would expect that $y_0'=y_0$. However, in our
calculations of surface tensions and coverages in slits (see Fig.~\ref{slit-fig}) we have
seen that for {\em very} small slit widths ($l<\delta$) $\rho^{\rm av}_{\rm 2d} \to 0$.
The limiting distance $\delta$ can be estimated from  the small $l$ expansion of
the contact density in slits, eq.~(\ref{small_l}):  
\bea
 \delta(\rho) \approx \rho^{-1}\exp(-\mu^{\rm ex}) \;.
\eea
This is indeed a  small length compared to $\sigma$: $\delta(0.5)\approx 4\cdot 10^{-2} $,
$\delta(0.7) \approx 1\cdot 10^{-3}$. But the depleted area in the annular wedge up to
distances $\delta$ must be added to the cavity, and therefore
\bea
 \label{y0p}
 y_0'\approx \sqrt{ R'( (1-z)+\delta)} \;.
\eea
  
The problem of the insertion energy of an additional cavity was the starting point of scaled
particle theory; here we can use the two--dimensional version \cite{Hel61} to obtain
\bea
 \label{spt}
 \mu_{\rm cav}(y_0') &=& \left\{ \begin{matrix} 
  p_{\rm 2d} \; \pi y_0'^2 + \gamma_{\rm 2d}\;2\pi y_0' + \epsilon_{\rm 2d}  & \qquad (y_0' > 1/2) \\
   -\log( 1-\pi \rho_{\rm 2d} y_0'^2 ) & \qquad (y_0' < 1/2) \end{matrix} \right. \\
  p_{\rm 2d} &=& \frac{4}{\pi} \frac{\eta_{\rm 2d}}{(1-\eta_{\rm 2d})^2} \; , \\
  \gamma_{\rm 2d} &=& -\frac{2}{\pi}  \frac{\eta_{\rm 2d}^2}{(1-\eta_{\rm 2d})^2} \;  \\
  \epsilon_{\rm 2d} &=& -\eta_{\rm 2d}\frac{ 1-2\eta_{\rm 2d}}{(1-\eta_{\rm 2d})^2} 
      -\log(1-\eta_{\rm 2d}) \\
 && \qquad \left(\eta_{\rm 2d} = \frac{\pi}{4} \rho_{\rm 2d}\right) \; .
\eea
The contact to the original surface energy of the slit, eq.~(\ref{om_sur}), is made by setting
$\gamma(l_0) = -p_{\rm 2d}$, thus it follows that $\sigma(y_0) \to \gamma_{\rm 2d}$. To obtain 
numbers,
we simply choose $l_0$ such that $\gamma(l_0)=2\gamma_\infty$. 
Using eq.~(\ref{g3d}) for the 3D surface
tension, we can determine $\rho_{\rm 2d}$ as a function of $\rho$, the 3D density of small
spheres. Remember that physically it would be also quite sensible to identify the 2D density
via eq.~(\ref{coverage}), $\rho_{\rm 2d} =  \rho^{\rm av}_{\rm 2d}$. 
A quick glance at Fig.~\ref{slit-fig}
assures us that the two definitions of $\rho_{\rm 2d}$ are quite consistent with each 
other\footnote{From $2\gamma_\infty(\rho) = -p_{\rm 2d}(\rho_{\rm 2d})$ it follows that
$\rho_{\rm 2d}= 0.33$ for $\rho=0.5$ and $\rho_{\rm 2d}= 0.59$ for $\rho=0.7$.}.

One might be concerned that the validity of eq.~(\ref{spt}) is limited for intermediate disk sizes
$y_0'=1 \dots2$ (where we would need it if calculating numbers for $\alpha=10$, say) 
as scaled particle theory is 
by construction only an interpolation between the
known analytical behavior of $ \mu_{\rm cav}(y_0')$ for $y_0'<1/2$ on the one side and 
$y_0' \to \infty$ on the other side. In view of lack of appropriate data in the literature
we have performed a quick MC check for two densities $\rho_{\rm 2d}=0.4$ and 0.6, the results
are shown in appendix \ref{mc-app}. From these results it follows that scaled particle theory
is precise enough for our purposes.

The final result for the depletion force, following from eqs.~(\ref{e1},\ref{e2}) and the 
considerations of the previous paragraphs, takes the form:
\bea
 \frac{F_\alpha(z<1)}{\pi R'}  &=&  \frac{1}{\pi R'}\left(p+\frac{2\gamma_\infty}{R'}\right)\;\frac{dV_s}{dz} -
 \frac{1}{\pi R'}\frac{d\mu_{\rm cav}}{dz} \;, \\    
 \label{force-model}
         &=&  \left(p+\frac{2\gamma_\infty}{R'}\right)\left( (z-1) + \frac{(z-1)^2}{4R'} \right)
 + \left\{ \begin{matrix} \rho_{\rm 2d}\frac{\di 1-\frac{1-z+\delta}{2R'}}{\di 1-\pi\rho_{\rm 2d} y_0'^2} &
   \qquad (y_0' < 1/2) \\ \\ \left(p_{\rm 2d} + \frac{\di \gamma_{\rm 2d}}{\di y_0'}\right)
  \left( 1-\frac{\di 1-z+\delta}{\di 2R'} \right) & \qquad (y_0' > 1/2) \end{matrix} \right. \\
   y_0' &=& \sqrt{ (1-z+\delta)\left(R'-\frac{1-z+\delta}{4}\right)}  \\
\eea
Here, for the sake of completeness, the exact geometrical expression for the cavity 
radius $y_0'$ is given. The expression
in eq.~(\ref{y0p}) is the leading term in an expansion of $y_0'$ with respect to $R'$.

Most remarkably, it follows for the force at $z=1$ (to first order in $1/R'$),
\bea
 \label{F-contact}
 \frac{F_\alpha(z=1)}{\pi R'}  &=&  \left\{ \begin{matrix} 
 \frac{\di \rho_{\rm 2d}}{\di 1 -\pi \delta R' \rho_{\rm 2d}} & \qquad (\delta R' < 1/4) \\ \\
   -2\gamma_\infty + \frac{\di \gamma_{\rm 2d}}{\di \sqrt{\delta R'}} & \qquad (\delta R' > 1/4) \end{matrix} \right. \; .
\eea

\section{Results from the annular slit approximation and comparison with DFT and MD}

\begin{figure}
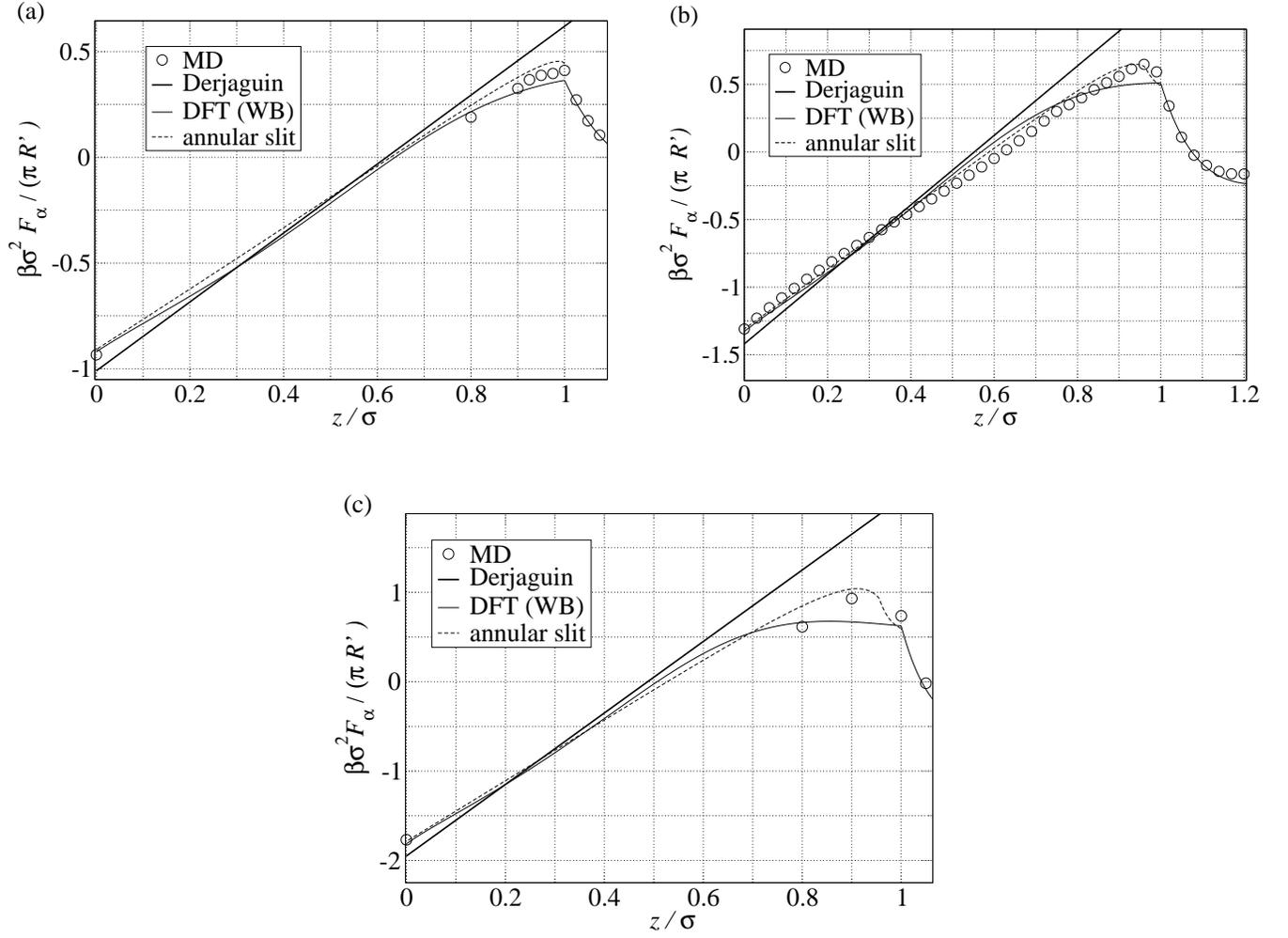

 \begin{center}
  \epsfig{file=fig4a.eps, width=8.5cm} \hspace{5mm}
  \epsfig{file=fig4b.eps, width=8.5cm} \\[9mm]
  \epsfig{file=fig4c.eps, width=8.5cm}
 \end{center}
 \caption{ The scaled force between two colloidal particles in the depletion region 
 $z<1$ for a size ratio $\alpha=10$ and for three solvent densities: (a) $\rho=0.5$, (b) $\rho=0.6$ 
 and (c) $\rho=0.7$. 
 Comparison between
 the annular slit approximation (eq.~(\ref{force-model})), 
  MD data \cite{Bib96} and test particle consistent DFT based on the White Bear (WB) functional. }
 \label{F1-fig}
\end{figure}

\begin{table}
 \begin{center}
  \begin{tabular}{p{0.8cm}p{0.8cm}p{1.3cm}p{1.2cm}p{1.2cm}p{1.2cm}p{1.2cm}} \hline \hline \\
    $\rho$ & $\rho_{\rm 2d}$ & $\delta$ &  \multicolumn{4}{c}{$\frac{\di F_\alpha(z=1)}{\di \pi R'}$} \\ \\
    & & & eq.~(\ref{F-contact}) & MD & DFT & Derjag. \\ \\ \hline \\
   0.4 & 0.22 & 0.18 & 0.28 & 0.22 & 0.22 & 0.31 \\
   0.5 & 0.34 & 0.04 & 0.44 & 0.41 & 0.36 & 0.62 \\
   0.6 & 0.46 & $8\cdot 10^{-3}$ & 0.49 & 0.57 & 0.51 & 1.15 \\
   0.7 & 0.59 & $1\cdot 10^{-3}$ & 0.60 & 0.74 & 0.62 & 2.05 \\
   0.8 & 0.71 & $5\cdot 10^{-5}$ & 0.71 &      & 0.64 & 3.55 \\
   0.9 & 0.81 & $8\cdot 10^{-7}$ & 0.81 &      & 0.62 & 6.01 \\ \\ \hline \hline 
  \end{tabular}
 \end{center}
  \caption{Results for the scaled depletion force at $z=1$, $F_\alpha/(\pi R')$, for the 
   annular slit approximation, from MD simulations \cite{Bib96}, test particle 
   consistent DFT and the Derjaguin approximation ($\alpha=10$).
   Note that the annular slit approximation predicts that the value of the scaled force at 
   this point is essentially
   given by $\rho_{\rm 2d}$, the effective 2D density.  }
  \label{F-tab}
\end{table}

Having obtained a closed expression for the force in the depletion region, eq.~(\ref{force-model}), we
can compare results to the available MD data and to the DFT results. Instead of using
the results from Ref.~\cite{Rot00} we apply the bridge functional formalism of
Ref.~\cite{Ros93} to obtain results which are ``one test particle consistent''. 
In effect, the equations for the depletion potential are transformed into RHNC type equations
where Rosenfeld's density functional (or extensions thereof) is the generating source for
the bridge diagrams. 
For more details we
refer to App.~\ref{dft-app} where we have outlined the procedure and compared to the previous 
DFT results.
Summarizing the results from the appendix,  the self--consistency for one test particle 
gives a relatively small shift of the depletion force for $z<0.6\dots 0.7$ which is always
upwards. This adds up to a 10 \dots 20 \% correction upwards for the depletion potential
at contact, $W_\alpha(z=0)$. For $z>0.7$, the results are  {\em quantitatively almost unchanged}.
Especially the failure of the previous results to converge to the Derjaguin limit
remains unaltered.

For the largest ratio $\alpha=10$ where simulation data are available, we show results for 
the depletion force in Fig.~\ref{F1-fig}. Values for the force at $z=1$ are compared in 
Tab.~\ref{F-tab}. In general, the agreement between the simulations and our annular slit
approximation is surprisingly
good. The approximation follows the trend of the simulation data to produce a maximum in the
depletion force for $z < 1$ and $\rho>0.5$. A pronounced maximum is absent in the
DFT results for the densities shown. Despite the better agreement of the annular slit
approximation with the simulation 
data, it is hard to tell whether for this size ratio $\alpha$ there is already a serious problem with
DFT. First, there are no error bar estimates for the MD data available, and secondly, the 
approximation
suffers from possible errors due to a finite number of  particles in our idealized annular slit.
This number can be estimated by
\bea
  N_s = \rho_{\rm 2d} A_{\rm wed} \approx \rho_{\rm 2d} \pi R' l_0
\eea
Indeed, since $l_0 \approx 0.2 \dots 0.4$ (see Fig.~\ref{slit-fig}) $N_s <5$ for all densities!
According to this estimate, our considerations should become increasingly reliable for
larger $\rho_{\rm 2d}$ and larger $\alpha$. Better correspondence with the simulation results
with increasing $\rho_{\rm 2d}$ is indeed observed, see Fig.~\ref{F1-fig}. 

\begin{figure}
 \begin{center}
  \epsfig{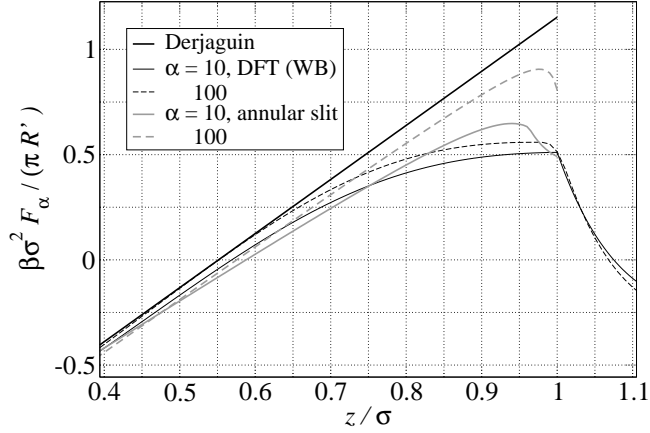}
 \end{center}
 \caption{Scaled depletion force for size ratios $\alpha=10$ and $\alpha=100$: Comparison between
  the annular slit approximation (gray curves) and test particle improved DFT based on the
  White Bear (WB) functional. The solvent density is $\rho=0.6$. 
  Note that the annular slit approximation approaches
  the Derjaguin limit quite slowly, nevertheless the Derjaguin limit is not reached at all
  by the DFT results.}
 \label{F2-fig}
\end{figure}

For larger $\alpha$, the discrepancy between the annular slit approximation and the DFT 
results becomes striking. We show
this in Fig.~\ref{F2-fig} for two size ratios, $\alpha=10$ and $\alpha=100$. As we have explained, the
annular slit approximation can be expected to become more accurate for larger $\alpha$ 
and it has the correct limiting 
behavior, so the conclusion would be that DFT becomes increasingly unreliable for $\alpha>10$.
Although not shown in the figure, there is already a substantial difference for $\alpha=20$, say.
Thus one should regard with extreme caution the claim of \cite{Rot00} 
that insertion route DFT can be expected to be rather accurate also for size ratios larger than 10.   
A similar claim made about a bridge diagram improved HNC treatment of the depletion potential
(see Ref.~\cite{Dic97}) should also be treated with caution as the HNC results show similar 
defects as the DFT results. Recall that the improved, test particle cosistent DFT results shown 
here can be viewed as 
HNC results with bridge diagram corrections supplied by the density functional and both
methods can be formulated in the language of insertion route DFT.

Finally we calculate the quantity $W_\alpha(0)-W_\alpha(1)$ (for $\alpha=10$)  which is 
roughly the depletion potential difference
between colloid contact and the first minimum for medium to high densities. 
The results are collected in Tab.~\ref{dw-tab}.
We see that the previous DFT results predict that the potential at colloid contact is minimal
for all values of $\rho$.
This finding is not changed by imposing test particle consistency; only the absolute value
of the potential difference is reduced  somewhat. The annular slit approximation 
predicts that the absolute 
minimum jumps to $z \approx 1$ for $\rho \approx 0.83$, still far away from the Derjaguin value
0.68. For $\alpha=100$, however, the jump of the absolute minimum occurs at a density of 0.73,
 according to the annular slit approximation.

\begin{table}
 \begin{center}
  \begin{tabular}{rrrrr} \hline \hline \\
    $\rho$ &  \multicolumn{4}{c}{$W_\alpha(0)-W_\alpha(1)$} \\ \\
    &          & test particle & annular slit \\
    &  DFT \cite{Rot00} & improved DFT  & approximation & Derjaguin \\ \\ \hline
   0.4 & --3.65 & --3.50 & --3.15 & --3.30 \\
   0.5 & --4.32 & --4.01 & --3.46 & --3.35 \\
   0.6 & --4.69 & --4.16 & --3.50 & --2.27 \\
   0.7 & --4.70 & --3.84 & --2.89 &   0.87 \\
   0.8 & --4.39 & --3.19 & --1.02 &   7.79 \\
   0.9 & --3.77 & --2.38 &   3.12 &  21.64 \\ \\ \hline \hline
  \end{tabular}
 \end{center}
  \caption{ Depletion potential difference $W_\alpha(0)-W_\alpha(1)=\int_0^1F_\alpha(z)dz$ 
  ($\alpha=10$) calculated using insertion route
  DFT \cite{Rot00}, test particle improved DFT based on the White Bear functional, 
  the annular slit approximation and the 
  Derjaguin approximation.}
  \label{dw-tab}
\end{table}

\section{Summary and conclusions}

In this paper, we have analyzed the depletion force between hard colloids in  a solvent consisting 
of small hard spheres. 
We started from an already previously observed disagreement between  simulation data/results 
from Rosenfeld's DFT and the Derjaguin limit. Albeit first derived on purely phenomenological
grounds, equilibrium statistical mechanics strongly supports the validity of the Derjaguin
limit for large size ratios $\alpha$ between colloids and solvent spheres.
The disagreement  between  simulation data/DFT
and the Derjaguin limit in the depletion zone for $\alpha=10$ could be explained by an 
effective approximation
which analyzes the structure of the solvent between the two colloids in terms of a (fairly)
dense 2D gas of hard disks. 
The depletion force near the onset of the depletion zone
(i.e. where exactly one solvent sphere fits between the colloids) is mainly determined by the force
to create a disk cavity in the effective 2D gas. Requiring the Derjaguin limit
for $\alpha\to \infty$, there are no free  parameters for the 2D gas. 
The agreement with simulation data
is very good even at the relatively small size ratio $\alpha=10$. 

For higher size ratios $\alpha>10$ no simulation data are available, and existing DFT results
showed no convergence towards the Derjaguin limit. Imposing test particle consistency we
calculated improved DFT results which however did not alter their large $\alpha$ behavior. 
Already for $\alpha=20$ the disagreement of the DFT depletion force with the results of the 
effective annular slit approximation becomes pronounced. We conclude that the limit of reliability
of insertion route DFT is reached for $\alpha=10$.
Through the test particle consistent calculations we have shown that insertion route DFT
and integral equation approaches are methodologically equivalent to each other;
the single variants differ in their choice for the bridge functional. Similar limits
for the reliablility can therefore be also expected for integral equations.
Likewise a similar limit will apply
if one treats solvophobic colloids in attractive fluids with repulsive cores using
Rosenfeld's DFT for hard sphere reference systems. To reach larger size ratios, the analysis 
of the annular slit approximation could also be extended to this case.

We have shown that quite subtle packing effects between the colloids play a role in determining
the depletion force for medium to large size ratios. In light of this it is actually amazing
that insertion route DFT (which explicitly needs only the density distribution around
{\em one} colloid) captures most of the effects and only misses the intricate effect of the
quasi--2D gas. We emphasize again that the insertion procedure is formally exact but
we possess only approximate expressions for the hard sphere density functional whose functional
derivative is needed for the insertion procedure to work.
The two variants investigated herein, the Rosenfeld and the White Bear functional, 
are -- despite being very precise -- not exact. One deficiency, if not the main, 
lies in the bulk direct correlation
functions $c^{(n)}$ of order $n\ge 2$: they are zero outside the hard core according
to the functionals, but from simulations and integral equations we know otherwise.   
Requiring test particle consistency has fixed this shortcoming for $c^{(2)}$, but for
all higher order ($n>2$) correlation functions one should require consistency for 
$n-1$ fixed test particles. Part of the problem for $n=3$ is thus the determination
of the density profile with the two colloids fixed. 
Therefore, it would be interesting to see how brute force DFT fares for size ratios
$\alpha>10$, i.e. whether one could observe the oscillatory packing in the annular wedge
directly and how it is related to $c^{(3)}$.

Throughout the paper we have argued that the Derjaguin limit for the depletion force is
meaningful. The annular slit approximation, eq.~(\ref{force-model}), predicts the leading correction
to the force $\propto \alpha^{-1/2}$. Interesting enough, this is a non--analytic term but
the Derjaguin limit is still reached continuously in the variable $1/\alpha$. 
Since the depletion force is also
connected to  an integral over the  surface densities, eq.~(\ref{F-mc}), this constitutes a hint
that the density profile also contains non--analytic contributions in $1/\alpha$. 
In fact, whereas there are good arguments that the density profile around hard convex objects 
should have an analytic
expansion in terms of the curvatures \cite{Rot03a}, such an analyticity requirement does not hold
for profiles around non--convex objects (such as the two--colloid configuration in the
depletion region). Consequently, there is the possibility that for $1/\alpha \to 0 $ 
the surface densities do not reach the Derjaguin limit of the surface densities
in a planar slit configuration, eq.~(\ref{small_l}). Possible singular contributions 
can not be understood with the current theories due to entropic arguments. 
A finite difference between the surface density and its Derjaguin limit would also
constitute a surprise since it would point to small sphere correlations 
which are much larger than the bulk correlation length and allude to a mysterious phase transition.
Nevertheless the non--analyticity of density profiles around curved objects is an extremely
interestig subject in itself which is currently being explored \cite{Rot03b}.

\section*{ACKNOWLEDGMENTS}
The author wants to thank R.~Evans, R.~Roth and S.~Dietrich for
enlightening discussions on the subject and a careful reading of the manuscript.

\begin{appendix}

\section{Quality of scaled particle theory for hard disks in 2D}
\label{mc-app}

\begin{figure}
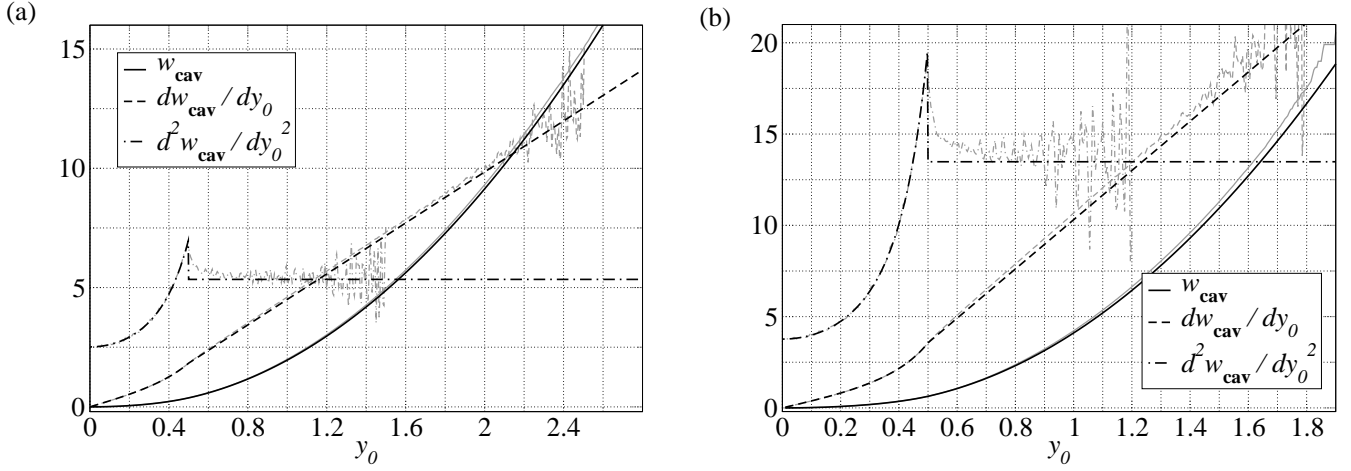

 \begin{center}
  \epsfig{file=fig6a.eps, width=8.5cm} \hspace{5mm}
  \epsfig{file=fig6b.eps, width=8.5cm}
 \end{center}
 \caption{Monte Carlo results for the insertion energy (and its first two derivatives) 
 of an (exclusion) disk of radius $y_0$
 into a system of hard disks with diameter 1. The thick black lines are the scaled particle 
  predictions, the rugged gray lines are raw data. The disk densities are: (a) $\rho_{\rm 2d}=0.4$
  and (b)  $\rho_{\rm 2d}=0.6$.}
 \label{mc-fig}
\end{figure}

To check the reliability of the 2D scaled particle theory predictions for the insertion energy
of a cavity with radius $y_0$, $w_{\rm cav}(y_0)$, we performed Monte Carlo tests for two medium 
densities, $\rho_{\rm 2d}=0.4$ and $\rho_{\rm 2d}=0.6$. The insertion energy is given
by
\bea
 w_{\rm cav}(y_0) = \log p_0(y_0)\;,
\eea
where $p_0(y_0)$ is te probability to find no center of a disk within a circle of radius
$y_0$ around an arbitrary point. Therefore we chose for each Monte Carlo move a new, 
random point, around
which we checked the latter condition \cite{Ada74}. The results, obtained for 4000 disks and
roughly $10^8$ moves, are depicted in Fig.~\ref{mc-fig}. The MC results are compared to 
scaled particle theory, and it is found that its simple prediction 
\bea
  w_{\rm cav}(y_0)= \pi y_0^2\;p_{\rm 2d} + 2\pi y_0\; \gamma_{\rm 2d}
\eea
with $\gamma_{\rm 2d}$ {\em independent} on the cavity radius is extremely good except in the
vicinity of $y_0=1/2$. At this point, exact analysis \cite{Hel61} demands that the third derivative
has a singularity\footnote{In Ref.~\cite{Hel61}, this singularity is given as $\propto (y_0-1/2)^{-1}$
which must be a misprint.},
\bea
  \frac{d^3w_{\rm cav}}{dy_0^3}(y_0\to 1/2_+) \propto (y_0-1/2)^{-1/2} \;.
\eea
This leads to a square root like cusp in the second derivative as can be seen in the MC results.
Scaled particle theory ignores this cusp which is not too bad an approximation since the cusp 
quickly relaxes
to a constant which is the pressure of the disk system. Apart from this effect of the non-analyticity
of $w_{\rm cav}$, scaled particle theory is sufficiently precise for our purposes.

\section{Test--particle consistent DFT}
\label{dft-app}

In this section we shortly explain the route taken by \cite{Rot00} 
for obtaining the depletion potential, and introduce corrections thereto by requiring
test particle consistency. Numerical results for both methods are presented.
Throughout this section $\beta=\sigma=1$.

Suppose we have a mixture of two hard species with particle radii $R_1$ and $R_2$. Species
2 refers to the colloids and the mutual interaction potentials are given by $u_{ij}(r)$. 
We split the density functional describing the mixture into 
an ideal gas and an excess part,
\bea
 {\cal F}[\{\rho_i(\vect r)\}] &=& {\cal F}_{\rm id}[\{\rho_i(\vect r)\}] + 
\  {\cal F}_{\rm ex}[\{\rho_i(\vect r)\}] \; .
\eea
The hierarchy of direct correlation functions is defined as
\bea
 \label{dcf-n}
  c^{(n)}_{i_1\dots i_n}(\vect x_1 \dots \vect x_n) = - \frac{ \delta^{(n)} {\cal F}_{\rm ex}}
     {\delta \rho_{i_1}(\vect x_1) \dots \delta \rho_{i_n}(\vect x_n)} \; .
\eea
Specifically for $n=1$  one can introduce an excess chemical potential functional,
\bea
 \mu^{\rm ex}_i[\vect x;\{\rho_i(\vect r)\}] = -c^{(1)}_i(\vect x)\; ,
\eea
which reduces to the usual excess chemical potential if the densities are constant,
$\mu^{\rm ex}_i[\vect x;\{\rho_i(\vect r)={\rm const}\}] = \mu^{\rm ex}_i(\{\rho_{i,0}\})$.

Suppose we have an inhomogeneous situation where one colloid is fixed. The depletion potential
is then defined as the difference between the work needed to put another colloid particle 
into the system at position $\vect x$ on the one hand and at infinity on the other hand.
Using the potential distribution theorem \cite{Hen83} we find
\bea
 \label{dep1}
 W_\alpha(x) = \lim_{\rho_2\to 0} \left( \mu^{\rm ex}_2[\vect x;\{\rho_i(\vect r)\}] - 
  \mu^{\rm ex}_2(\{\rho_{i,0}\})
    \right)\; ,
\eea
where we have assumed that $\lim_{|\vect x|\to\infty}\rho_1(\vect x) = \rho_{1,0}$. Note that 
$\mu^{\rm ex}_2[\dots]$ depends in the required limit only on the density distribution
$\rho_1(r)$ of species 1 around the colloid, i.e. {\em before} the second colloid is
inserted. With a given excess functional at hand, the depletion potential is found by
obtaining $\rho_1(r)$ through grand potential minimization,
\bea
 \label{gp-min}
  0 &=& \frac{\delta F}{\delta \rho_i(\vect x)} - \mu_i + V_i(\vect x)  \qquad \to \\
  -\log(\rho_1(\vect x)) &=& \mu^{\rm ex}_1[\vect x;\{\rho_1(\vect x'),0\}] - 
  \mu_1^{\rm ex}(\{\rho_{1,0},0\}) + u_{12}(\vect x) \;,
\eea
and then inserting this solution into eq.~(\ref{dep1}).

\subsection{Test particle consistency}

The depletion potential is the negative potential of mean force between a pair
of colloids at infinite dilution. The following relation is valid:
\bea
  W_\alpha(|\vect x|) = -\log g_{22}(|\vect x|;\{\rho_{1,0},0\})  - u_{22}(|\vect x|) \; ,
\eea 
where $g_{22}$ is the colloid--colloid distribution function in the limit of vanishing
colloid density. On the one hand, this distribution function can be determined via the
depletion potential described in the manner above. On the other hand, the excess functional
defines the second--order correlation function $c^{(2)}_{ij}$ 
through eq.~(\ref{dcf-n}) (for $n=2$) which
in turn can be inverted to give $g_{ij}$ using the Ornstein--Zernike relation
\bea
 h_{ij}(|\vect r|) - c^{(2)}_{ij}(|\vect r|) &=& \sum_k \rho_{k,0}\; h_{ik}\ast c^{(2)}_{kj} 
 (|\vect r|)\; , \\
    h_{ij}(|\vect r|) &=& g_{ij}(|\vect r| ) -1 \;, \\
  h_{ik}\ast c^{(2)}_{kj} (|\vect r|) &=& \spaceint{}{\vect r'}  h_{ik}(|\vect r'|) 
  c^{(2)}_{kj}(|\vect r-\vect r'|) \; .  
\eea
In general, both routes will give different results for an approximated free--energy functional.

To make both routes consistent with each other, we proceed as follows \cite{Ros93}: Consider
the equation which relates the bridge function $b_{ij}$ to the distribution and direct correlation function,
\bea
 \label{bridgedef}
  b_{ij}(|\vect r|) &=& -\log g_{ij}(|\vect r|) - u_{ij}(|\vect r|) + \sum_k \rho_{k,0}\; 
 h_{ik}\ast c^{(2)}_{kj}(|\vect r|) \; .
\eea
The various closures of integral equations follow by specifying a model for the bridge function,
e.g. $b_{ij}(r)=0$ for the HNC closure. The bridge function can be generated by a bridge 
functional which we define to be the functional which contains all contributions beyond second
order in a density expansion of the exact free energy functional around fixed, constant bulk 
densities
$\rho_{i,0}=\rho_i(\vect r)-\Delta\rho_i(\vect r)$:
\bea
 \label{fbr-def}
 {\cal F}[\{\rho_i(|\vect r|)\}] &=& {\cal F}_{\rm id}[\{\rho_k(|\vect r|)\}] +  
 F_{\rm ex}(\{\rho_{k,0}\}) +
    \mu^{\rm ex}_i (\{\rho_{k,0}\}) \spaceint{}{\vect r}\Delta\rho_i(\vect r)  - \\
   &&   \frac{1}{2}\spaceint{}{\vect r}\spaceint{}{\vect r'} c^{(2)}_{ij}(\vect r,\vect r';{\rm bulk})
  \Delta\rho_i(\vect r) \Delta\rho_j(\vect r') +
    {\cal F}_{\rm ex}^{\rm br}[\{\rho_k(\vect r)\}] \; .
\eea
Doubly occuring indices are summed over. To verify that the such introduced 
${\cal F}_{\rm ex}^{\rm br}$ indeed generates the bridge functions, we minimize the grand potential
according to eq.~(\ref{gp-min}) in the presence of the interparticle potential, $V_i=u_{ij}$.
\bea
  \frac{\delta {\cal F}_{\rm ex}^{\rm br}}{\delta \rho_i(\vect r)} &=&
   -\log \frac{\rho_i(\vect r)}{\rho_{i,0}} - u_{ij}(\vect r) + 
   c_{ik}^{\rm(2)} \ast \Delta\rho_k (\vect r) \; .
\eea
Since $\rho_{i,0}\; g_{ij}(|\vect r|) = \rho_i(|\vect r|)$ and 
$\rho_{i,0}\; h_{ij}(|\vect r|)=\Delta\rho_i (|\vect r| ) $ we have recovered
eq.~(\ref{bridgedef}) upon the identification 
\bea
  b_{ij}(\vect r) = \left.\frac{\delta {\cal F}_{\rm ex}^{\rm br}}{\delta \rho_i(\vect r )}
   \right|_{V_i=u_{ij}} \;.
\eea 
Up to now everything was exact but in order to specify a closure explicitly we assert that
the true bridge functional can be approximated by the bridge functional of a reference model
for which we possess an explicit form of ${\cal F}_{\rm ex}$:
\bea
  \label{fbr-approx}
   {\cal F}_{\rm ex}^{\rm br}[\{\rho_k(\vect r)\}]  \approx  
 {\cal F}_{\rm ex}^{\rm ref}[\{\rho_k(\vect r)\}] - 
  \mu^{\rm ex, ref}_i (\{\rho_{k,0}\}) \spaceint{}{\vect r}\Delta\rho_i(\vect r) + 
   \frac{1}{2}\spaceint{}{\vect r}\spaceint{}{\vect r'} 
 c^{(2), {\rm ref}}_{ij}(\vect r,\vect r';{\rm bulk})\Delta\rho_i(\vect r) \Delta\rho_j(\vect r') \; .
\eea
A remark is in order here. Note  that eqs.~(\ref{fbr-def},\ref{fbr-approx}) together define
a new functional which is now test particle consistent, i.e. 
the inversion of the Ornstein--Zernike relation gives the same result as an explicit determination
of the distribution functions through the density profiles around test particles.
This consistency holds regardless of the form of the interparticle potential and of
how good or bad the choice of the reference system is. 
Of course, the reference system of choice is again hard spheres described by Rosenfeld's functional
or a recently improved version, the White Bear functional \cite{Rot02}. 

\begin{figure}
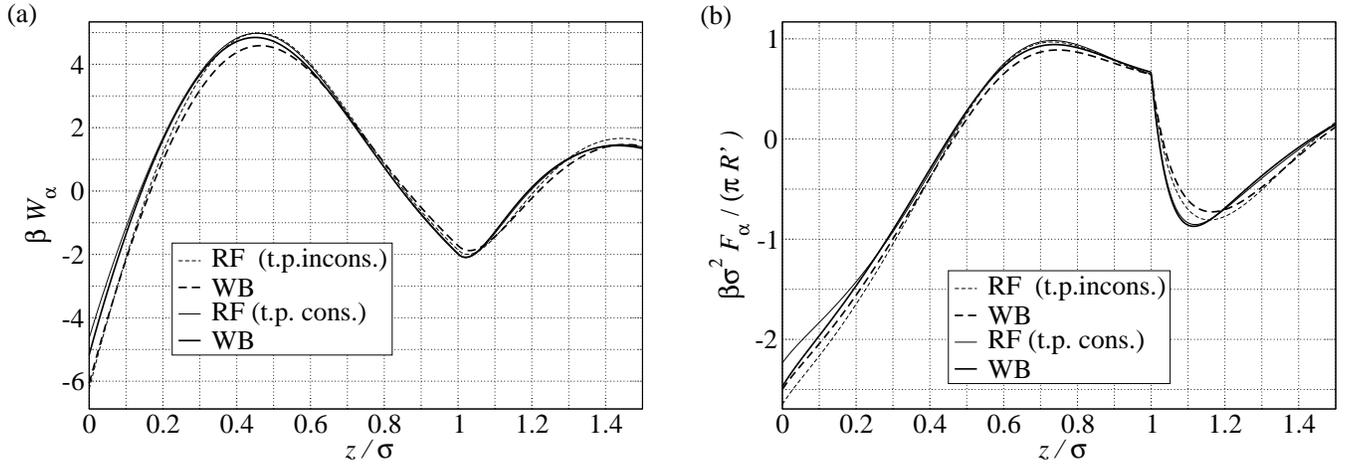

 \begin{center}
  \epsfig{file=fig7a.eps, width=8.5cm} \hspace{5mm}
  \epsfig{file=fig7b.eps, width=8.5cm}
 \end{center}
 \caption{(a) Depletion potential and (b) depletion force between two colloids for a solvent density of 
  $\rho=0.8$ and a size ratio
  $\alpha=10$: Comparison between test particle inconsistent and consistent results, 
  using Rosenfeld's (RF)  and
  the White Bear (WB) functional respectively.}
 \label{dep1-fig}
\end{figure}

In the limit $\rho_2 \to 0$, relevant for the determination of the depletion potential, the equations
for the distribution functions decouple. For $g_{11}$ we have to solve the following equations
($r=|\vect r|$):
\bea
 h_{11}(r) - c^{(2)}_{11}(r) &=&  \rho_{1,0}\; h_{11}\ast c^{(2)}_{11} (r)\; , \\
 -\log g_{11}(r) -  u_{11}(r) &=& \rho_{1,0}\; h_{11}\ast (c^{(2), {\rm ref}}_{11}-c^{(2)}_{11}) (r) + 
 \left( \mu^{\rm ex,ref}_1[r;\{\rho_{1,0}\;g_{11}(r),0\}] - \mu^{\rm ex,ref}_1(\{\rho_{1,0},0\}) \right) \;.\quad
\eea
As expected, the colloids decouple and we are left with the equations for the one--component
system.
Employing the White Bear functional,
we obtain one--component distribution functions which fit the MC data even better than
the standard Verlet parametrization \cite{Ver72,Ver72a}.
The input of $h_{11},\; c_{11}$ is needed to solve 
the next two equations for $g_{12}$ (the normalized density distribution around one colloid):
\bea
 h_{12}(r) - c^{(2)}_{12}(r) &=&  \rho_{1,0}\; h_{11}\ast c^{(2)}_{12} (r)\; , \\
 -\log g_{12}(r) -  u_{12}(r) &=& \rho_{1,0}\; h_{12}\ast (c^{(2), {\rm ref}}_{11}-c^{(2)}_{11}) (r) +
 \left( \mu^{\rm ex,ref}_1[r;\{\rho_{1,0}\;g_{12}(r),0\}] - \mu^{\rm ex,ref}_1(\{\rho_{1,0},0\}) \right)\;. \quad
\eea
Having obtained $h_{12},\; c_{12}$, the depletion potential is simply given by
\bea
 \label{dep2}
  W_\alpha(r)=-\log g_{22}(r) -u_{22}(r) = \rho_{1,0}\; h_{12}\ast (c^{(2), {\rm ref}}_{12}-c^{(2)}_{12}) (r) 
+ \left( \mu^{\rm ex,ref}_2[r;\{\rho_{1,0}\;g_{12}(r),0\}] - \mu^{\rm ex,ref}_2(\{\rho_{1,0},0\}) \right)\;.
\eea

Comparing the expressions for the depletion potential for test particle inconsistent  and
consistent DFT, eqs.~(\ref{dep1},\ref{dep2}), we see that the main difference is 
buried in the first term on the rhs of eq.~(\ref{dep2}) since $\rho_1(r)|_{\rm inconsistent}
\approx \rho_{1,0}\;g_{12}(r)|_{\rm consistent}$ .

Results for the size ratio  $\alpha=10$ reveal no huge differences between the 
test particle consistent
and inconsistent calculations. For distances between the colloids $z<0.6\dots 0.7$ the consistent
results give a somewhat higher force than the inconsistent results which adds up to a noticeable
upward shift in $W_\alpha(z=0)$. 
Apart from that the differences are minimal, even at higher densities.
For $\rho_{1,0}=0.8$, we show the depletion potential and force in Fig.~\ref{dep1-fig}, calculated
with the Rosenfeld and the White Bear functional.

 Due to the accuracy in the one--component case, this choice of reference
system has been extended to binary soft systems \cite{Kah96}. Again, the agreement with simulation 
data is extremely good but the size ratio in the binary systems was well below 10. 
Only recently the depletion potential between soft colloids in soft fluids has been 
calculated for size ratios of about 10 using this method \cite{Amo01}. In general, test particle
consistent DFT fares much better than any other theoretical method compared to the
simulation data. Nevertheless, in the case of hard colloids in a LJ fluid discrepancies
to the simulation data occur which show the same footprints as the deviations we observe here
in the case of hard colloids in hard fluids.

\end{appendix}


\begin{thebibliography}{99}
\bibitem{Ver72a}
There is a small error in Ref.~\cite{Ver72}.
The first order estimate for the decay length of the oscillations, eq.~(2.17) in the paper, 
should be changed:\\
\centerline{$
  \mu\; d\approx \frac{\di 24A/d}{\di \eta_W g_W(1,\eta_W)} \to \frac{\di 24A/d}{\di \eta g(1,\eta)}\;. 
$}
The second order correction becomes important for higher densities:\\
\centerline{$
 \mu\; d\approx \frac{\di 24A/d}{\di \eta g(1,\eta)+4(d_W^3-4+3/d_W)g'(1,\eta)}\;.
$}
The error also appears in the widely used Ref.~\cite{Han76}.
\end{thebibliography}
\end{document}